\documentclass[final,5p,times,twocolumn,numbers,sort&compress]{elsarticle}

\usepackage{blindtext}

\usepackage{graphicx}
\usepackage{dcolumn}
\usepackage{bm}
\usepackage[utf8]{inputenc}
\usepackage{amsmath}
\usepackage{amsfonts}
\usepackage{mathtools}
\usepackage{amssymb}
\usepackage{frontespizio}
\usepackage[T1]{fontenc}
\usepackage[utf8]{inputenc}
\usepackage{lmodern}
\usepackage[english]{babel}
\usepackage{latexsym}
\usepackage{braket}
\usepackage{physics}
\usepackage{slashed}
\usepackage{hyperref}
\usepackage{verbatim}
\usepackage{graphicx}
\usepackage{units}
\usepackage{booktabs}
\usepackage{appendix}

\usepackage{color}
\def\laq{~\raise 0.4ex\hbox{$<$}\kern -0.8em\lower 0.62ex\hbox{$\sim$}~}
\def\gaq{~\raise 0.4ex\hbox{$>$}\kern -0.7em\lower 0.62ex\hbox{$\sim$}~}

\usepackage{titlesec}					
\usepackage{float}

\usepackage{comment}

\providecommand{\quotes}[1]{``#1''}

\def\beq{\begin{equation}}
	\def\eeq{\end{equation}}
\def\bea{\begin{eqnarray}}
	\def\eea{\end{eqnarray}}
\def \bes{
\begin{split}}
    \def \ees{\end{split}}

\journal{Physics Letters B}

\begin{document}

\begin{frontmatter}

\title{Tensor-Induced Backreaction in Ultra-Slow-Roll Inflation}

\author[a,b,c]{Alessandro Ciaiolo}
\ead{a.ciaiolo@studenti.unipi.it}

\author[a,b]{Giovanni Marozzi}
\ead{giovanni.marozzi@unipi.it}

\affiliation[a]{
    organization={Dipartimento di Fisica, Universit\`a di Pisa},
    addressline={Largo B. Pontecorvo 3},
    postcode={56127},
    city={Pisa},
    country={Italy}
}

\affiliation[b]{
    organization={Istituto Nazionale di Fisica Nucleare, Sezione di Pisa},
    addressline={Largo B. Pontecorvo 3},
    postcode={56127},
    city={Pisa},
    country={Italy}
}

\affiliation[c]{
    organization={PPGCosmo, Universidade Federal do Esp\'irito Santo},
    postcode={29075-910},
    city={Vit\'oria},
    state={ES},
    country={Brazil}
}

\begin{abstract}
We discuss the impact of tensor-induced backreaction at second order in perturbation theory on primordial black holes production in single-field inflationary models, where the comoving curvature power spectrum is amplified by a transient ultra-slow-roll phase. 
Although the leading perturbative correction is mostly negligible in standard slow-roll evolution, it becomes important in non-attractor regimes, going in the direction of limiting the suppression of the first slow-roll parameter $\varepsilon$ and reducing the enhancement of the scalar power spectrum. 
We then go beyond the standard perturbative analysis by considering the background parameters as effective quantities with evolution self consistently determined by a closed system of backreaction equations.
This analysis, which leads to a partial resummation of the effect beyond the 
perturbative treatment, confirms the limitation of the suppression of $\varepsilon$ by consistently showing how the effect remains important 
even after accounting for the effective evolution of the background in the evaluation of the backreaction itself.
Finally, we discuss how, taking into account the tensor-induced backreaction, the fine-tuning associated with the value of the peak amplitude of the scalar power spectrum is strongly reduced.
\end{abstract}


\end{frontmatter}

\section{Introduction}
Originally proposed to overcome the shortcomings of the standard Hot Big-Bang theory \cite{Guth:1980zm,Linde:1981mu}, the inflationary paradigm, already in its minimal version with only one field, is also sufficient to furnish a causal mechanism to generate nearly scale invariant spectra of scalar and tensor fluctuations \cite{Mukhanov:1981xt,Starobinsky:1979ty}, 
in particular for the observationally important scales associated to CMB and large-scale structure formation \cite{Planck:2018jri}.  
Within this single-field inflationary models class, the models that undergo 
an ultra-slow-roll (USR) regime also provide an efficient mechanism to amplify primordial curvature perturbations on scales related to the production of primordial black holes (PBHs), and not directly constrained by large-scale observations \cite{Kinney:2005vj,Martin:2012pe, Dimopoulos:2017ged}.

Inflationary phases are conveniently characterized through the Hubble hierarchy of slow-roll parameters,
\begin{equation}
\epsilon_{n+1} =\frac{d\ln|\epsilon_n|}{dN} \quad \mathrm{with} \quad \epsilon_0 = \frac{1}{H},
\label{parametri}
\end{equation}
where $H$ is the Hubble factor and $d N=H dt$ defines the number of e-folds.
In what follows, we adopt the notation
$\epsilon_1=\varepsilon$ and $\epsilon_2=\eta$. The standard slow-roll (SR) regime requires all slow-roll parameters to be small, whereas USR is a non-attractor phase in which the potential becomes extremely flat, the inflaton velocity rapidly decreases, and $\eta \simeq -6$. Hence, $\varepsilon \propto e^{-6N}$, allowing a large enhancement of the comoving curvature perturbation.
A common realization of this mechanism can be obtained in single-field scenarios with a transitional nearly flat region in the potential \cite{Garcia-Bellido:2017mdw,Di:2017ndc,LISACosmologyWorkingGroup:2025vdz,Mishra:2019pzq}. In these models, the inflaton, initially in a SR regime, enters a USR phase where the enhancement of small-scale perturbations occurs before returning to a final SR phase that ends inflation. The resulting enhancement of 
the comoving curvature perturbation may seed overdense regions that subsequently collaps into PBHs when these modes re-enter the Hubble horizon.
The abundance of PBHs depends exponentially on the amplitude of the scalar power spectrum, which typically must reach a value $A_\mathcal{R}^{\rm peak}\sim 10^{-3}$ -- $10^{-2}$ depending on the width and shape of the peak \cite{Gow:2020bzo}. 
Furthermore, such peak value is extremely sensitive to the details of the dynamics near the flat region. This generates what is typically considered a fine-tuning problem \cite{Cole:2023wyx, Frolovsky:2023hqd,Geller:2022nkr} (but see also \cite{Iovino:2025tcv}).

In this Letter we investigate how non-linear effects, arising from the second-order coupling between scalar and tensor perturbations, impact the background dynamics described above. Using a covariant and gauge invariant approach \cite{Gasperini:2009wp,Gasperini:2009mu}, we derive the leading tensor-induced backreaction to the effective Hubble rate and slow-roll parameters in a generic single-field model, going beyond the results obtained in \cite{Marozzi:2014xma} for standard SR,  and discussing the backreaction enhancement in non-attractor regimes. We then go beyond the standard perturbative treatment, where the perturbations live on a fixed background. 
Namely, we use a self-consistent approach that takes into consideration the effective evolution of the background in the evaluation of the backreaction itself.
Finally, as an illustrative example, we apply the above results to a benchmark potential \cite{Mishra:2019pzq,Cole:2023wyx} for PBHs production, obtaining that the tensor induced backreaction strongly suppresses the enhancement of the scalar power spectrum, by limiting the suppression of $\varepsilon$. 
Furthermore, we show how also the fine-tuning associated with $A_\mathcal{R}^{\rm peak}$ is strongly reduced.
We use $c=1$, while $M_{Pl}^2=1/(8\pi G)$.

\section{Gauge Invariant approach to the quantum backreaction}
The homogeneous and isotropic description of cosmology should be regarded as an effective large-scale approximation rather than the exact geometry of the Universe. 
Therefore, metric and matter inhomogeneities of the true spacetime must be averaged to recover an effective homogeneous description,
giving rise to additional backreaction terms \cite{Buchert:1999er,Buchert:2001sa}. 
This issue is particularly relevant during inflation, when indeed quantum perturbations of the inflaton coupled to the metric perturbations are stretched beyond the Hubble scale and may strongly affect the effective evolution. 

The impact of quantum backreaction during the inflationary era has a long and controversial history
(see, for example, 
\cite{
  Tsamis:1992sx,
  Tsamis:1993ub,
  Tsamis:1996qm,
  Tsamis:1996qq,
  Tsamis:1996qk,
  Tsamis:2013cka,
  Brandenberger:2002sk,Mukhanov:1996ak,Abramo:1997hu,
  Unruh:1998ic, Abramo:1998hj, 
  Abramo:2001dc,
Abramo:2001db, 
   Geshnizjani:2002wp,
      Geshnizjani:2003cn,
      Losic:2005vg,
      Marozzi:2006ky}).
Among the main challenges there was the one of avoiding unphysical artifacts in the backreaction evaluation. In order to face this issue, in \cite{Finelli:2011cw} the averaged contribution of perturbations was then formulated in terms of gauge invariant observables.
This approach starts with the consideration that a backreaction effect is meaningful
only when “measured” by a given observer. 
Following \cite{Gasperini:2009wp,Gasperini:2009mu}, we
then introduce a scalar field $A(x)$, with timelike gradient, and define the averaging hypersurfaces
$\Sigma_{A_0}$, where $A(x)=A_0$. 
Setting a barred coordinate system $\bar{x}^{\mu}=(\bar{t},\mathbf{x})$, where $A$ is homogeneous, and defining $t_0$ as the value of the time coordinate $\bar t$ for which
$\bar{A}(\bar{x})=A^{(0)}(\bar{t})=A_0$,
the gauge invariant average of a scalar quantity $S(x)$ is then \cite{Gasperini:2009wp,Gasperini:2009mu}
\begin{equation}
\left\langle S \right\rangle_{A_0}
=
\frac{
\left\langle
\sqrt{|\bar{\gamma}(t_0,\mathbf{x})|} \bar{S}(t_0,\mathbf{x})
\right\rangle
}{
\left\langle
\sqrt{|\bar{\gamma}(t_0,\mathbf{x})|}
\right\rangle
} ,
\end{equation}
where $\bar{\gamma}$ is the determinant of the induced three-dimensional metric on $\Sigma_{A_0}$.
The properties of the observers sitting on $A(x)=A_0$ are then univocally determined \cite{Marozzi:2010qz} adopting the foliation associated to the unit normal vector
$    n^\mu
=-Z_A^{-1/2}\partial^\mu A $,
with $Z_A=-\partial_\mu A\partial^\mu A$. 
Within this framework one can then identify the effective scale factor, that describes the space-time dynamical evolution 
as seen by the observers comoving with the hypersurface $\Sigma_{A_0}$, with
$a_{eff}=\langle \sqrt{|\bar{\gamma}|}
\rangle^{1/3}$.

The cosmological equation for the averaged geometry
that describe the effective Hubble parameter is then given by \cite{Gasperini:2009mu}:
\begin{equation}
H^2_{eff}=\left(\frac{1}{a_{eff}}\frac{\partial \, a_{eff}}{\partial A_0} \right)^2=\frac{1}{9}\left\langle
\frac{\Theta}{Z_A^{1/2}}\right\rangle_{A_0}^2,
\label{H^2eff}
\end{equation}
where $\Theta=\nabla_\mu n^\mu$ is the expansion scalar of the timelike congruence $n^\mu$.
Hereafter, we set the background of the scalar clock such as 
$A^{(0)}(t)=A_0=t$, so that the derivative wrt $A_0$ corresponds to the proper time derivative and 
standard results are recovered neglecting perturbations \cite{Marozzi:2010qz}.
From the definition of $H_{eff}$, we can define 
the effective number of 
e-folds by $d N_{eff}= H_{eff} d A_0$.
The non-perturbative effective description of the Hubble hierarchy of slow-roll parameters is then given by
\begin{equation}
\epsilon_{n+1,\, eff} =\frac{d\ln|\epsilon_{n,\, eff}|}{dN_{eff}} \quad \mathrm{with} \quad \epsilon_{0,\, eff} = \frac{1}{H_{eff}}\,,
\label{parametri-eff}
\end{equation}
from which one obtains the following results
\begin{equation}
  \varepsilon_{eff}=-\frac{\dot{H}_{eff}}{H^2_{eff}} \quad,\quad \eta_{eff}=\frac{\dot{\varepsilon}_{eff}}{\varepsilon_{eff} H_{eff}}\,,
  \label{Eff-Slow-roll-parameters}
\end{equation}
where the dot stands for the derivative wrt $A_0$ or, equivalently, wrt the proper time $t$. 

Let us now take a spatially flat Friedmann-Lema\^itre-Robertson-Walker (FLRW) background geometry and
expand the metric perturbatively up to second order, as follows 
\begin{equation}
\begin{aligned}
g_{00} &= -1 - 2\alpha - 2\alpha^{(2)}, \\
g_{0i} &= -\frac{a}{2}\left(\beta_{,i}+B_i\right)    -\frac{a}{2}\left(\beta^{(2)}_{,i}+B_i^{(2)}\right), \\
g_{ij} &= a^2 \Bigg[
\delta_{ij}\left(1-2\psi-2\psi^{(2)}\right)
+D_{ij}\left(E+E^{(2)}\right)  \\
&
\!\!\!+\frac{1}{2}\left(\chi_{i,j}+\chi_{j,i}+h_{ij}\right)
+\frac{1}{2}\left(\chi^{(2)}_{i,j}+\chi^{(2)}_{j,i}+h^{(2)}_{ij}\right)
\Bigg] \,,\label{GeneralGauge}
\end{aligned}
\end{equation}
where
$D_{ij}=\partial_i\partial_j-\delta_{ij}(\nabla^2/3)$, and
 first-order perturbations are written without an explicit
superscript. The quantities $\alpha$, $\beta$, $\psi$ and $E$ denote scalar perturbations,
whereas $B_i$ and $\chi_i$ are transverse vector perturbations ($\partial^i B_i=\partial^i\chi_i=0$). Finally,  $h_{ij}$ is a  transverse and traceless tensor
perturbation 
($\partial^i h_{ij}=0$ and $h^i{}_i=0$).

Considering a minimally coupled single-field inflation, the matter sector is described by the following action 
\begin{equation}
    S = \int d^4x \sqrt{-g} \left[ 
\frac{R}{16{\pi}G}
    - \frac{1}{2} g^{\mu \nu}
    \partial_{\mu} \Phi \partial_{\nu} \Phi - V(\Phi) \right] ,
    \label{action}
\end{equation}
where the inflaton field can be decomposed as $\Phi(x)=\phi(t)+\varphi(x)+\varphi^{(2)}(x)$.

Let us now give the second-order perturbative expansion 
of Eq. (\ref{H^2eff})
and of the effective slow-roll parameters of Eqs. (\ref{Eff-Slow-roll-parameters}). 
 Such perturbative expressions will then describe the space-time dynamic, as seen from an observer sitting on the $\Sigma_{A_0}$ hypersurface, at the leading perturbative order.

 Furthermore, we limit ourselves to computing corrections to homogeneous values due to long wavelength (LW) quantum fluctuations, we can then use the LW limit in our computations.
In such LW limit one then gets \cite{Finelli:2011cw}

\begin{align}
\begin{split}
\bar{\Theta} &= 3H
- 3H\bar{\alpha}
- 3\dot{\bar{\psi}}
+ \frac{9}{2}H\bar{\alpha}^2
+ 3\bar{\alpha}\dot{\bar{\psi}}
- 6\bar{\psi}\dot{\bar{\psi}} \\
&\quad - 3H\bar{\alpha}^{(2)}
- 3\dot{\bar{\psi}}^{(2)}
- \frac{1}{8} h_{ij}\dot{h}^{ij}\,,
\end{split} \\
\bar{Z}_A &= 
1 - 2\bar{\alpha}
+ 4\bar{\alpha}^2
- 2\bar{\alpha}^{(2)} \,, \\
\sqrt{|\bar{\gamma}|}
&= a^3 \left(
1 - 3\bar{\psi}
+ \frac{3}{2}\bar{\psi}^2
- \frac{1}{16} h_{ij} h^{ij}
- 3\bar{\psi}^{(2)}
\right).
\end{align}
Then, simply substituting these expressions into Eqs. (\ref{H^2eff}) and (\ref{Eff-Slow-roll-parameters}), we obtain
\begin{align}
H_{eff}^{2} &= H^{2} + 2 H B, \quad\quad \varepsilon_{eff} = \varepsilon - \frac{2\varepsilon}{H}B - \frac{1}{H^2}\dot{B}\,, 
\label{H^2perturbazioni} \\
\eta_{eff} &= \eta - \frac{2}{H}\left(\varepsilon+\frac{\eta}{2}\right)B - \frac{1}{H^2}\left(4-\frac{\eta}{\varepsilon}\right)\dot{B}-\frac{\ddot B}{H^3\varepsilon}\,, 
\label{etaperturbazioni}
\end{align}
where $B=\langle \dot{\bar{\psi}}\bar{\psi} \rangle
- \langle \dot{\bar{\psi}}^{(2)} \rangle
- \frac{1}{24}\langle \dot{h}^{ij}h_{ij} \rangle$.
The formalism described above, with regard to the evaluation of the expansion rate, 
has already been extensively applied to the standard single-field SR regime by considering different observers \cite{Marozzi:2013uva,Marozzi:2011zb,Marozzi:2012tp,Galanti:2024jhw,Marozzi:2014xma}.  
However, the system of Eqs. (\ref{H^2perturbazioni})-(\ref{etaperturbazioni}) is valid at the second perturbative order in the LW approximation, but for any value of $\varepsilon$ and $\eta$.
We then have a set of equations that are suitable for investigating the backreaction in any dynamical regimes. 
In particular, in the following we will analyze the tensor-induced backreaction in generic models, extending the results obtained in \cite{Marozzi:2014xma}, where a standard chaotic potential was considered.

\section{Tensor-induced backreaction}

During inflation perturbations originate from quantum fluctuations and, once stretched beyond the Hubble radius, can be then effectively described as classical stochastic perturbations.
To evaluate the backreaction of such inhomogeneities on the background evolution we
have now to choose the observer wrt such backreaction
has to be evaluated.
In our context we have a physically natural choice, since, in single-field inflation, the inflaton is the only dynamical matter field 
that can be used as a clock.
Hence, we consider a comoving observer who sees a homogeneous inflaton field for which $\varphi=\varphi^{(2)}=0$.

The metric decomposition introduced in (\ref{GeneralGauge}) contains redundant degrees of freedom, that have to be removed by a gauge choice. 
Fixing the gauge typically removes two scalar and one vector perturbations at each order. The key case for us is
the uniform field gauge (UFG), which is indeed defined by setting $\Phi(x)=\phi(t)$ (i.e. $\varphi=\varphi^{(2)}=0$) and 
by another condition (one can consider $g_{i0}=0$).

Following \cite{Marozzi:2014xma}, 
we then note that 
at the second perturbative order fluctuations becames each other coupled. In particular,
every second order scalar contribution can be decomposed as $\delta^{(2)}=\delta^{(2)}_s+\delta^{(2)}_t$,  where $\delta^{(2)}_s$ is the purely scalar dependent part, while $\delta^{(2)}_t$ is the part induced by the presence of first order tensor modes.
We neglect the contribution that would be induced by the presence of first order vector modes, because these are kinematically suppressed during inflation. 
Hereafter we will focus ourself only on the tensor induced backreaction, neglecting the purely scalar one. A discussion regarding this scalar backreaction and the reason why we chose to neglect it in the present work can be found in Appendix A. 

 By including only pure and induced tensor contributions,
 we obtain that the $B$ contribution 
 becomes
\begin{equation}
    B=- \langle \dot{\bar{\psi}}_t^{(2)} \rangle
- \frac{1}{24}\langle \dot{h}^{ij}h_{ij} \rangle\,.
\label{Btensor}
\end{equation} 
Therefore, to evaluate 
Eqs. (\ref{H^2perturbazioni})-(\ref{etaperturbazioni}) it is sufficient to evaluate the correlators $\langle \dot{\bar\psi}^{(2)}_t\rangle$ and $\langle h_{ij}\dot h^{ij}\rangle$. The procedure for obtaining these correlators is 
described in Appendix B, where we obtain 
\begin{equation}
\langle\dot h_{ij} h^{ij} \rangle
=\frac{4}{\pi^2}\frac{H^3}{M_{Pl}^2} \quad, \quad \langle \dot{\bar\psi}^{(2)}_t \rangle
=
\frac{13}{16\pi^2 }
\frac{H^3}{M_{Pl}^2\varepsilon} \,.
\label{corr}
\end{equation}

By using Eqs. (\ref{corr})  in 
Eqs. (\ref{H^2perturbazioni})-(\ref{etaperturbazioni}), and performing the calculation at leading order in the first slow-roll parameter, but considering that in non-attractor regimes $\eta=\mathcal{O}(1)$, after some algebra we obtain 
\begin{equation}
\begin{aligned}
H_{eff}^2 &= H^2 \left[ 1 - \frac{13}{8\pi^2} \frac{H^2}{M_{Pl}^2\varepsilon}\right], \\
\varepsilon_{eff} &= \varepsilon \left[ 1 - \frac{13}{16\pi^2} \frac{H^2}{M_{Pl}^2\varepsilon} \left( \frac{\eta}{\varepsilon} + 1 \right) \right], \\
\eta_{eff} &= \eta \left[ 1 + \frac{13}{16\pi^2} \frac{H^2}{M_{Pl}^2\eta} \left( 2 + \frac{8\eta}{39} + \frac{4\eta}{\varepsilon} + \frac{2\eta^2}{\varepsilon^2} - \frac{\dot{\eta}}{H \varepsilon^2} \right) \right].
\end{aligned}
\label{eff-results}
\end{equation}
The above corrections are suppressed by the prefactor $H^2/M_{ Pl}^2$ and remain under perturbative control in standard SR, in agreement with what was found in \cite{Marozzi:2014xma}. On the other hand, while the relative correction to $H_{ eff}^2$ scales always as $1/\varepsilon$, in USR regime the induced relative correction to $\varepsilon_{ eff}$ and $\eta_{ eff}$ scale as $1/\varepsilon^2$.
Hence, in models undergoing a USR phase, the obtained perturbative backreaction corrections to $\varepsilon_{ eff}$ and $\eta_{ eff}$ become non-perturbative when $\varepsilon=\mathcal{O}\left(\frac{H}{M_{Pl}}\right)$.

\section{Beyond the perturbative approach for ultra-slow-roll}

The corrections appearing in Eqs. (\ref{eff-results}) are obtained perturbatively on a fixed background. In order to estimate the effect of the backreaction on the background when such backreaction has a non-negligible impact, we start by evaluating the backreaction on the background Hubble parameter taking in consideration the corrected background in the evaluation of such backreaction. We then determine the corresponding number of e-folds and slow-roll parameters in a consistent way.
This leads to the 
following self-consistent procedure, which is well 
defined as long as the magnitude of the correction on $H^2$ is perturbative, i.e. $|\delta H^2(H,\varepsilon)| \ll H^2$. 
The procedure consists of keeping the functional form of $\delta H^2$, but  determining it using the corrected effective background quantities. Namely, we define a self-consistent effective Hubble parameter $H_{sc}$ as
follows
\begin{equation}
H_{{sc}}^2
= H^2-
\frac{13}{8\pi^2}
\frac{H^4_{sc}}{M_{Pl}^2\varepsilon_{sc}} \,,
\label{HscFirst}
\end{equation}
where we now
obtain a self-consistent definition for the slow-roll parameters by deriving $H_{sc}$ as follows
\begin{equation}
  \varepsilon_{sc}
=-\frac{d\ln H_{sc}}{d N_{sc}}\quad,
\quad
\eta_{sc}
= \frac{d\ln \varepsilon_{sc}}{d N_{sc}} \,,
\label{eq:self_consistent_closure}  
\end{equation}
where we have introduced the definition $dN_{sc}=H_{sc}dt$. 
By considering the corrected background in the evaluation of the backreaction we obtain a partial resummation of the effect,
reliable as long as the condition $|\delta H^2(H,\varepsilon)| \ll H^2$ holds.

To numerically solve the closed system, given by Eqs. (\ref{HscFirst}) and (\ref{eq:self_consistent_closure}), 
we proceed as follows. 
Starting from Eqs. (\ref{eq:self_consistent_closure}) we can first write the following identity
\begin{equation}
    \frac{dH^2_{sc}}{dN}=-2\frac{H_{sc}}{H}\varepsilon_{sc} H^2_{sc}\,,
    \label{sistemasc1}
\end{equation}
then by differentiating Eq. (\ref{HscFirst}) wrt $N$ one obtains
\begin{equation}
\!\!\!\!\frac{d\varepsilon_{sc}}{dN}\!
=\!\frac{H_{sc}}{H}
\varepsilon_{ sc}^2\left[
\frac{16\pi^2}{13}\frac{M_{Pl}^2}{H_{ sc}^2}
\left(
\varepsilon
\left(
\frac{H}{H_{sc}}
\right)^3
-
\varepsilon_{sc}
\right)
-4
\right].
\label{sistemasc2}
\end{equation}
Finally, the equation for $\eta_{sc}$
is trivially derived from (\ref{sistemasc2}) as
\begin{equation}
    \eta_{sc}=\varepsilon_{ sc}\left[
\frac{16\pi^2}{13}\frac{M_{Pl}^2}{H_{ sc}^2}
\left(
\varepsilon
\left(
\frac{H}{H_{sc}}
\right)^3
-
\varepsilon_{sc}
\right)
-4
\right].
\label{sistemasc3}
\end{equation}
We finally numerically solve for $H_{\rm sc}^{2}$ and $\varepsilon_{sc}$ using Eqs. \eqref{sistemasc1} and \eqref{sistemasc2}, while $\eta_{sc}$ is derived from Eq. \eqref{sistemasc3}.
The self-consistent quantities obtained as functions of $N$ are then plotted against $N_{sc}$. 
The initial conditions used
are fixed by matching 
$H_{sc}$ with $H_{eff}$ at the 
initial time
$N_0=0$, i.e
$
H_{sc}^2(N_0)
=
H_{eff}^2(N_0)
$.
Once this matching condition is imposed, the initial values of $\varepsilon_{sc}$ and $\eta_{sc}$
 are 
 determined by Eqs. (\ref{HscFirst}) and (\ref{sistemasc3}) at $N_0$.

Let us now assess the impact of tensor backreaction through the procedure introduced above on a benchmark model originally proposed in \cite{Mishra:2019pzq} and subsequently studied, for instance, in \cite{Cole:2023wyx}:
\begin{equation}
V(\phi)=V_0\frac{\phi^n}{\phi^n + M^n}\left[1+A \exp\left(-\frac{(\phi-\phi_d)^2}{2\sigma^2}\right)\right].
\end{equation}
In this model, the parameters $n$, $M$, and $V_0$ determine, almost independently of the remaining parameters, the agreement with large scale structures and CMB observations. A Gaussian bump is then added, controlled by the parameters $A$, $\phi_d$, and $\sigma$, which finely modify the dynamics near the quasi-flat region. We set $n=2$ and $M=M_{Pl}/2$, while $V_0$ is fixed by normalizing the scalar power spectrum at the CMB pivot scale. The latter is specified by the condition $\phi_*(N_*)=3\,M_{Pl}$, where the normalization $A^*_\mathcal{R}=2.1\times10^{-9}$ is imposed. With this choice, at the pivot scale one obtains $n_s=0.9648$ and $r=0.0026$, consistently with the observational bounds
(see, for example, \cite{BICEP:2021xfz,Planck:2018jri}) 
for any Gaussian-bump parameters combination we will consider. 
The initial value of the inflaton is set in the standard SR regime at $\phi_0=4\,M_{Pl}$. We fix $A=1.17\times10^{-3}$ and $\sigma=1.59\times10^{-2}\,M_{Pl}$, then the choice of $\phi_d$, 
to which value the model is strongly sensitive \cite{Cole:2023wyx}, 
fixes the position of the Gaussian bump, and distinguishes the two realizations of the model analyzed in the following. The first realization is defined by $\phi_d=\phi_{d1}=2.18801\,M_{Pl}$. The second one by $\phi_d=\phi_{d2}=2.18801\times1.00006\,M_{Pl}$.
\begin{figure}[h]
    \centering    \includegraphics[width=\columnwidth]{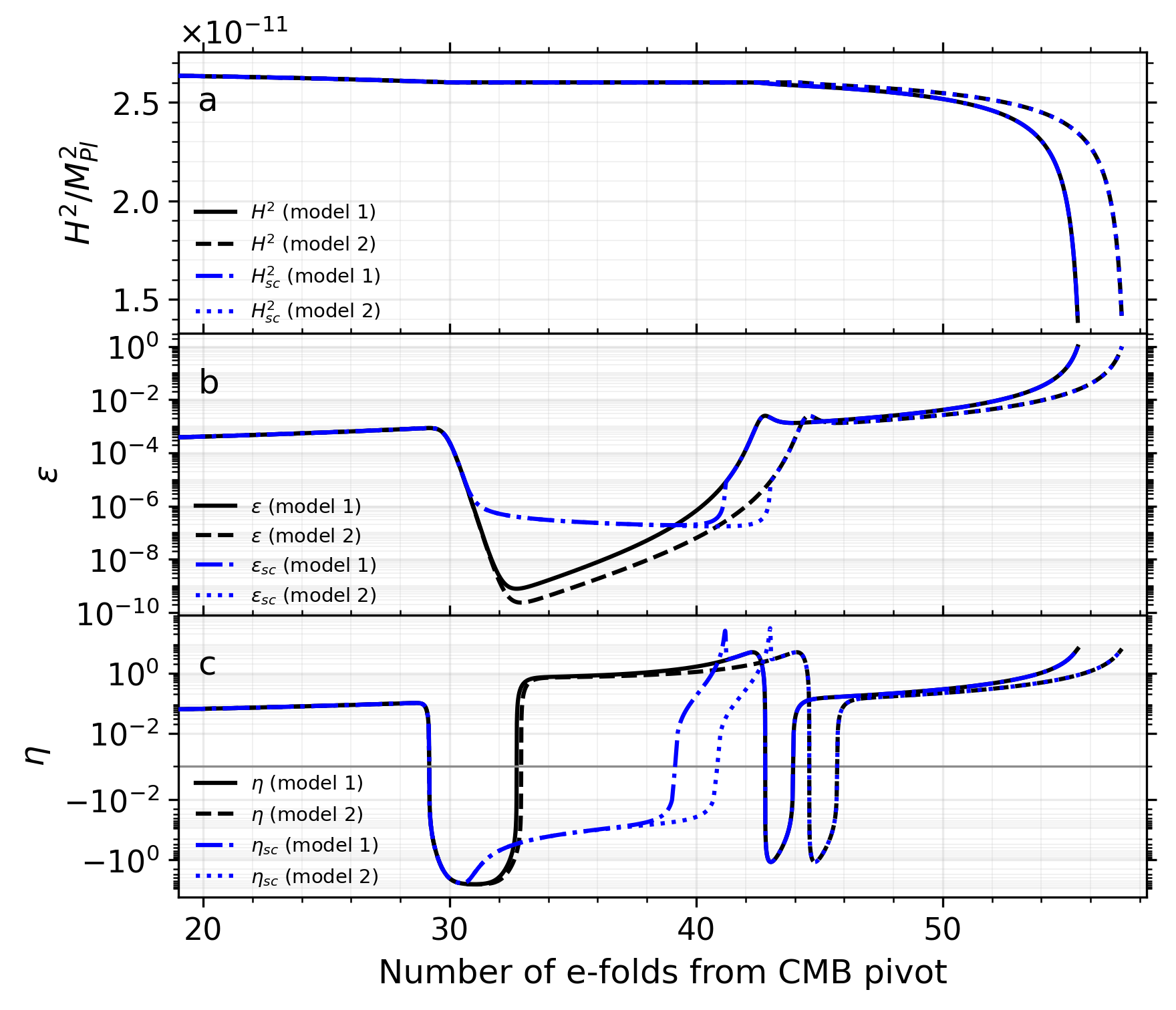}
    \caption{Evolution of $H^2$ and $H^2_{sc}$ in panel (a), of $\varepsilon$ and $\varepsilon_{sc}$ in panel (b), and of $\eta$ and $\eta_{sc}$ in panel (c). 
    Solid black curves are the first background model realization, and dot-dashed blue curves are the  corresponding self-consistent quantities.
    Dashed black curves are the second background model realization, 
    and dotted blue curves are the corresponding self-consistent quantities.
 }
    \label{fig:analisi}
\end{figure}

As it can be seen from panel (b) of Fig. \ref{fig:analisi}, in the first background model one reaches
$\varepsilon_{min\,1}\simeq 7.8\times 10^{-10}$, whereas in the second one has $\varepsilon_{min\,2}\simeq 2.3\times 10^{-10}$ and, in both cases, this occurs at a value of $H^2\simeq2.6\times 10^{-11} \,M_{Pl}^2$. 
From Eqs. (\ref{eff-results}) we then have 

\begin{equation}
\left| \frac{\delta H^2}{H^2} \right|_{max} = \frac{13}{8\pi^2} \frac{H^2}{M_{Pl}^2\varepsilon}\Big|_{max} = \mathcal{O}(10^{-2})
\end{equation}
 for both cases. Therefore, the tensor backreaction induces a perturbatively controlled correction to $H^2_{sc}$, for both cases throughout their entire evolution, and our self-consistent procedure is under control. 
 
On the contrary, for $\varepsilon_{eff}$ and $\eta_{eff}$ the tensor backreaction 
would lead to the breakdown of the perturbative approach shortly after the onset of the non-attractor phase, which is why the effective quantities are not shown in Fig. \ref{fig:analisi}. The main effect of the backreaction is of limiting the suppression of the first slow-roll parameter in the non-attractor region. This effect is milder 
when the backreaction is evaluated taking into account the new effective background.
We then plot
$\varepsilon_{sc}$ and $\eta_{sc}$ in panels (b) and (c) of Fig. \ref{fig:analisi}, together with the slow-roll parameters evaluated without backreaction.
The non-attractor phase, defined for values of $\eta<-3$ without backreaction and $\eta_{sc}<-3$ with backreaction, lasts respectively about $2.27$ and $2.45$ e-folds $N$ for the first and second model realizations in the case without backreaction, while it lasts about $1.01$ e-folds $N_{sc}$ for both cases, when the effect of tensor backreaction is considered.

After the end of the non-attractor phase, the first slow-roll parameter grows
for the background cases while stay almost constant including backreaction.
Finally, the dynamic with and without backreaction converge to the same values, leading to the end of inflation.
\begin{figure}[h]
    \centering    \includegraphics[width=\columnwidth]{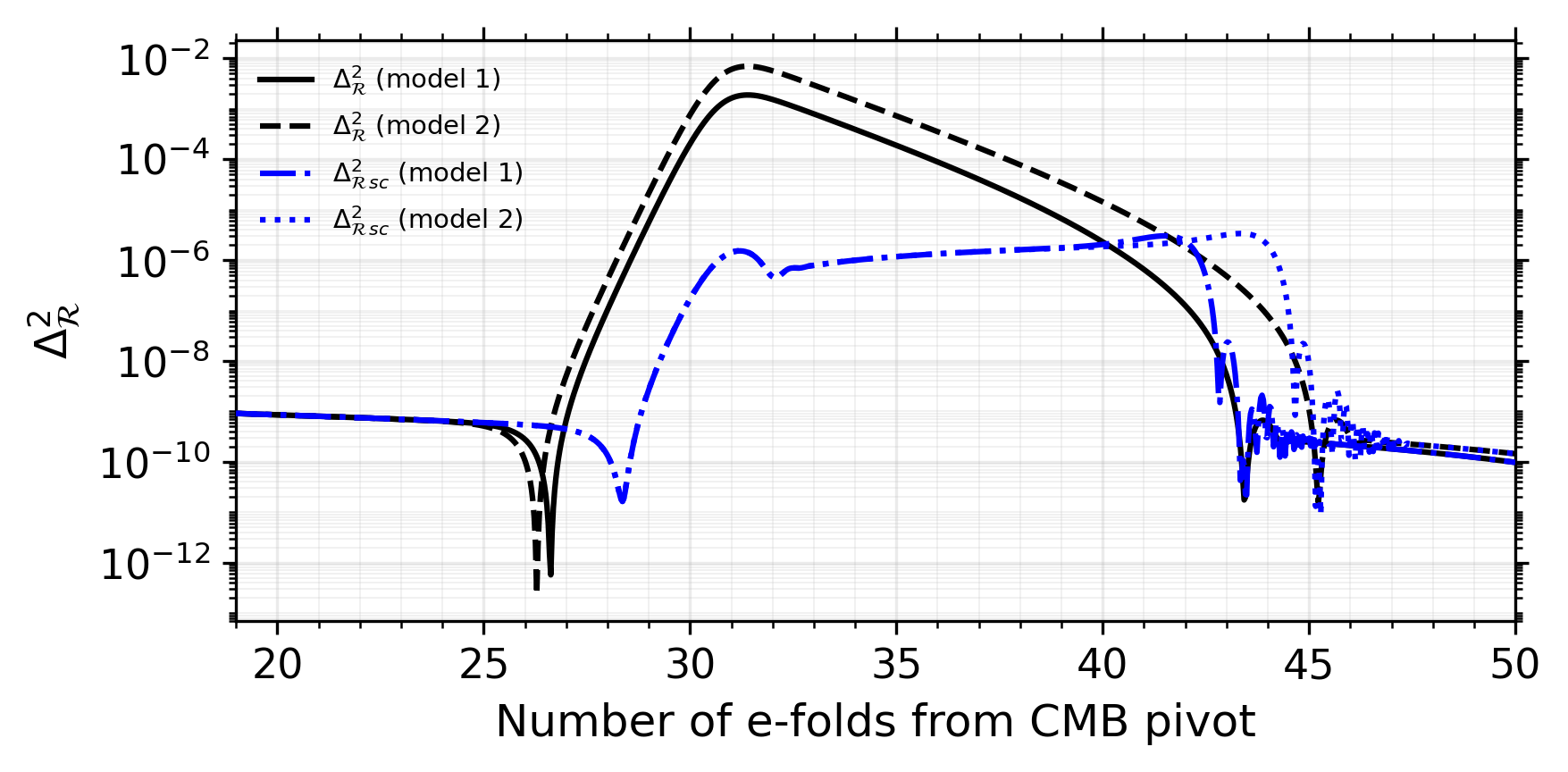}
    \caption{We show $\Delta_{\mathcal R}^2$ and $\Delta_{\mathcal R\,sc}^2$ in function of the e-folds for which the mode $k$ exit the horizon. 
    The solid black curve is $\Delta_{\mathcal R}^2$ for the first background model realization, and the dot-dashed blue curve is the corresponding $\Delta_{\mathcal R\,sc}^2$.
    The dashed black curve is $\Delta_{\mathcal R}^2$ for the second background model realization, 
    and the dotted blue curve is the corresponding $\Delta_{\mathcal R\,sc}^2$.}
    \label{fig:spettro}
\end{figure}

The scalar power spectra $\Delta_{\mathcal R}^2(k)=k^3|\mathcal R_k|^2/(2\pi^2)$ shown in Fig. \ref{fig:spettro} are obtained by solving the Mukhanov--Sasaki equation mode by mode for the comoving curvature perturbation,
\begin{equation}
\frac{d^2\mathcal R_k}{dN^2}
+\left(3-\varepsilon+\eta\right)\frac{d\mathcal R_k}{dN}
+\left(\frac{k}{a H}\right)^2\mathcal R_k=0,   
\end{equation}
with Bunch-Davies initial conditions imposed in the sub-Hubble regime. Each mode is evaluated at the end of inflation, where it is well frozen. $\Delta_{\mathcal{R}\, sc}^2$ is evaluated by replacing the background quantities with the corresponding self-consistent ones and by evolving the modes with respect to $N_{sc}$.
For both model realizations $\Delta_{\mathcal{R}\, sc}^2$ reaches a nearly equal maximum amplitude slightly below $2\times10^{-6}$ and stay nearly constant for a large range of e-folds. 
This should be compared with the corresponding cases without backreaction, where the spectrum reaches values about $2\times10^{-3}$ for the first model realization and $7\times10^{-3}$ for the second case. Tensor backreaction therefore plays a crucial role in assessing PBHs production, strongly suppressing it and changing the associated phenomenology in models with transient non-attractor phases.

Furthermore, let us put in evidence how  for both realizations $\Delta_{\mathcal{R}\, sc}^2$ reaches a nearly equal maximum amplitude, and the non-attractor phase lasts nearly the same number of e-folds, this shows that considering backreaction the model is much less sensitive to the details of the dynamics near the quasi-flat region of the potential.
A more quantitative analysis of how the USR fine-tuning problem \cite{Cole:2023wyx, Frolovsky:2023hqd,Geller:2022nkr} is alleviated by the backreaction effect is presented in Appendix C.

\section{Conclusions}
In this Letter we have shown how the production of PBHs is strongly modified if one takes into consideration the backreaction of tensor induced scalar perturbations.
Looking to Figs. \ref{fig:analisi} and \ref{fig:spettro}, we see how during an USR phase the backreaction suppresses the peak of the power-spectrum to nearly the same value and modifies the effective background dynamics. 
In particular, the duration of the non-attractor phase
is abbreviated and remains nearly constant in both cases when backreaction is considered.
So we have that tensor backreaction reduces the fine-tuning associated with the power-spectrum peak produced by a USR phase (see also the quantitative evaluation in Appendix C), but also gives a negligible PBHs production.

The above conclusions would suggest a no-go theorem for the production of PBHs within a USR phase of inflation, when one takes into account backreaction effects. However, before stating such a no-go theorem, one should be very careful. 
The proposed procedure for evaluating backreaction beyond standard perturbation theory provides a consistent partial resummation of the effects sourced by first-order tensor perturbations. However, it does not include genuine contributions from higher-order tensor perturbations. Moreover, the procedure has not been extended to the background inflaton equation of motion, and implicitly assumes that the functional form of the perturbation evolution equations remains unchanged. To conclude, our results clearly show how the contribution of backreaction is central for determining the true dynamics in USR scenarios, but, for a complete characterization of the actual phenomenology associated with these models, the extensions mentioned above and the actual evaluation of the scalar contribution should be investigated. We postpone this analysis to future works.\\

\section*{Acknowledgements}
We are very thankful to Chiara Animali and Pietro Conzinu for useful discussions and comments on a previous version of this manuscript.
We are supported in part by the Istituto Nazionale di Fisica Nucleare (INFN) through the Commissione Scientifica Nazionale 4 (CSN4),  under the Iniziativa Specifica (IS) Theoretical Astroparticle Physics (TAsP) and the IS Quantum Fields in Gravity, Cosmology and Black Holes (FLAG).
The work of AC was financed in part by the Coordenação de Aperfeiçoamento de Pessoal de Nível Superior - Brasil (CAPES) - Finance Code 001.
The work of GM was supported by the research grant number 2022E2J4RK \quotes{PANTHEON: Perspectives in Astroparticle and
Neutrino THEory with Old and New messengers} under the program PRIN 2022 funded by the Italian Ministero dell’Universit\`a e della Ricerca (MUR) and by the European Union – Next Generation EU.

\appendix

\section{Scalar backreaction for a comoving observer} Regarding the purely scalar contribution to cosmological backreaction in non-attractor regimes, several approaches have been explored in the literature, including Hartree, stochastic coarse-graining and lattice methods \cite{Cheng:2021lif,Caravano:2024moy,Geshnizjani:2004tf,Pattison:2019hef}. While, a perturbative, covariant, observer-dependent and gauge-invariant analysis based on averaged geometrical observables, along the lines of \cite{Marozzi:2014xma,Finelli:2011cw,Marozzi:2010qz,Marozzi:2013uva,Marozzi:2011zb,Marozzi:2012tp}, but extended to non-attractor regimes, is 
still missing.
On this regard in \cite{Marozzi:2013uva} (see also \cite{Finelli:2011cw}) was shown how in a  standard single field model of inflation, taking into consideration only the purely scalar backreaction, the comoving observers do not see any backreaction effect on the effective Hubble rate in the LW limit, for any potential and to all orders in the slow-roll parameters.
Let us then note that, for the class of comoving observers and in the LW limit, the perturbation $\bar{\psi}$ exactly coincides with the comoving curvature perturbation $\mathcal{R}$. Therefore, looking to Eqs. (\ref{H^2perturbazioni})-(\ref{etaperturbazioni}), we see how the scalar  contribution is associated to the dynamical evolution of $\mathcal{R}$ and is exactly zero when this is conserved on super-Hubble scales. This is indeed the case for standard  single field inflationary models, for which neglecting the tensor contribution 
the comoving curvature perturbation is conserved on super-Hubble scale. 
Under this prospective one can also more easily interpret the results obtained in \cite{Marozzi:2014xma} and \cite{Galanti:2024jhw}, where one obtains an induced non-zero backreaction for a comoving observer, respectively when it is considered the coupling between scalar and tensor at the second perturbative order and when the inflaton field is coupled to gauge fields in a model of $U(1)$ axion inflation. We indeed have that in these two cases the presence of another dynamical field coupled to the inflaton implies
the non-conservation of the comoving curvature perturbation on super-Hubble scale at the second perturbative order.

However, in the non-attractor regime, during the USR phase the comoving curvature perturbation is no longer conserved on the super-Hubble scales \cite{Martin:2012pe}.
Therefore, in these cases the proof of \cite{Marozzi:2013uva} does not apply.
Despite this,  the physical reason for which a comoving observer does not experience a purely scalar backreaction should hold also for non-attractor regimes. Such physical reason corresponds to the fact that the inflaton is the only matter field that sources the inflationary dynamics, so on the hypersurface where such a field is homogeneous there should be no dynamical scalar backreaction (see  also \cite{Geshnizjani:2002wp}).

Furthermore, looking to Figs. \ref{fig:analisi} and \ref{fig:spettro} the tensor backreaction both abbreviates the duration of the non-attractor phase, during which the comoving curvature perturbation is not conserved, and suppresses the amplitude of such comoving curvature perturbation. As a consequence, one would expect in any case a strongly reduced scalar backreaction. 

We let a more in-depth investigation of the above points for future works.

\section{Evaluation of $\langle \dot{\bar\psi}^{(2)}_t \rangle$ and $\langle\dot h_{ij} h^{ij} \rangle$} Let us first note that 
being our approach gauge invariant, we can perform our calculations in any gauge. We then chose to carry out them by fixing
the uniform curvature gauge (UCG), defined by
$g_{ij}=a^2\left[\delta_{ij}+\frac{1}{2} \left(h_{ij}+h^{(2)}_{ij}\right)\right]$.
Following 
\cite{Marozzi:2014xma}, under
a second-order coordinate transformation
\begin{equation}
x^\mu \rightarrow \bar{x}^\mu
=
x^\mu+\epsilon^\mu_{(1)}
+\frac{1}{2}
\left(
\epsilon^\nu_{(1)}\partial_\nu\epsilon^\mu_{(1)}
+\epsilon^\mu_{(2)}
\right)\,,
\end{equation}
we have the following tensor-induced changes of the scalar perturbations \cite{Marozzi:2010qz} 
\begin{align}
\bar{\psi}^{(2)}_t &= \psi^{(2)}_t + \frac{H}{2} \epsilon^0_{(2)} + \frac{1}{6} \nabla^2 \epsilon_{(2)}, 
\label{psitgauge} \\
\bar{\varphi}^{(2)}_t &= \varphi^{(2)}_t - \frac{\dot{\phi}}{2} \epsilon^0_{(2)} 
\quad,\quad 
\bar{\alpha}_t^{(2)} = \alpha_t^{(2)} - \dot{\epsilon}^0_{(2)}, \\
\bar{\beta}_t^{(2)} &= \beta_t^{(2)} - \frac{2}{a} \epsilon^0_{(2)} + 2 a \dot{\epsilon}_{(2)} 
\,,\, 
\bar{E}_t^{(2)} = E_t^{(2)} - \epsilon_{(2)},
\end{align}
where the second-order gauge generator is decomposed as
\begin{equation}
\epsilon^\mu_{(2)}
=
\left(
\epsilon^0_{(2)},
\partial^i\epsilon_{(2)}
+
\epsilon^i_{(2)}
\right)\quad,\quad
\partial_i\epsilon^i_{(2)}
=
0.
\end{equation}
As underlined in the main text the $\bar{x}^\mu$ reference frame is the UFG, then
starting from the UCG, we have 
\begin{equation}
\epsilon^0_{(2)}
=
2
\frac{\varphi^{(2)}_t}{\dot{\phi}}.
\end{equation}
By taking the average and in the LW limit we then obtain from Eq.(\ref{psitgauge})
\begin{equation}
\langle
\dot{\bar{\psi}}^{(2)}_t
\rangle
=
\left\langle
\frac{d}{dt}
\left(
\frac{H}{\dot{\phi}}
\varphi^{(2)}_t
\right)
\right\rangle.
\label{psidott}
\end{equation}
Following \cite{Finelli:2003bp, Marozzi:2014xma}, by combining the Hamiltonian and momentum constraints induced to second order by tensor we find
\begin{equation}
\frac{H}{a}
\nabla^2
\beta^{(2)}_t
=
\frac{\dot{\phi}^{\,2}}{H M_{Pl}^2}
\frac{d}{dt}
\left(
\frac{H}{\dot{\phi}}
\varphi^{(2)}_t
\right)
-
q_t
+
\frac{2V}{M_{Pl}^2}
s_t,
\label{laplacianobeta}
\end{equation}
where the tensor-induced source terms are
\begin{align}
\hspace{-1em}s_t
={}&
\frac{1}{8H}\frac{1}{\nabla^2}
\Bigg[
\frac{1}{2a}\beta^{,ij}\nabla^2 h_{ij}
+\alpha^{,ij}\dot h_{ij}
+\frac{1}{2}h^{km,i}\dot h_{mi,k}
\notag\\
&\hspace{1.2cm}
-\frac{1}{4}\dot h_{ij}\nabla^2 h^{ij}
-\frac{3}{4}h_{ij,k}\dot h^{ij,k}
-\frac{1}{2}h^{ij}\nabla^2\dot h_{ij}
\Bigg],
\label{st}
\\[0.5em]
\hspace{-1em}q_t
={}&
-\frac{H}{2a}\beta_{,ij}h^{ij}
+\frac{1}{8a}\beta_{,ij}\dot h^{ij}
-\frac{H}{4}h_{ij}\dot h^{ij}
-\frac{1}{32}\dot h_{ij}\dot h^{ij}
\notag\\
&\hspace{0.2cm}
+\frac{1}{8a^2}h^{ij}\nabla^2 h_{ij}
+\frac{3}{32a^2}h^{ij}{}_{,k}h_{ij}{}^{,k}
-\frac{1}{16a^2}h^{ij}{}_{,k}h_{ik}{}^{,j}.
\label{qt}
\end{align}
After taking the average, in the LW limit and at leading order in the first slow-roll parameter, the left-hand side of Eq. (\ref{laplacianobeta}) can be neglected, while it follows directly from Eq. (\ref{qt}) that
\begin{equation}
\langle
q_t
\rangle
=
-
\frac{H}{4}
\langle
h_{ij}
\dot{h}^{ij}
\rangle.
\label{qtfinal}
\end{equation}
To evaluate $\left\langle s_t\right\rangle$, we first note that the first two terms in Eq. (\ref{st}) do not contribute to the quantum average.
The remaining terms can be rearranged separating the expressions that are symmetric and antisymmetric under the
exchange of $h_{ij}$ and $\dot{h}_{ij}$. One then obtains
\begin{equation}
\langle
s_t
\rangle
=
\frac{1}{16H}
\langle
\frac{\partial^i\partial_k}{\nabla^2}
\left(
h^{km}
\dot{h}_{mi}
\right)
\rangle
-
\frac{3}{64H}
\langle
h_{ij}
\dot{h}^{ij}
\rangle.
\label{st2}
\end{equation}
The tensor perturbation can be expanded in Fourier modes as
\begin{equation}
h_{ij}(t,\mathbf{x})
=
\sum_s
\int
\frac{d^3k}{(2\pi)^{3/2}}
\left[
e_{ij}(\mathbf{k},s)
h_{s,k}(t)
a_{\mathbf{k},s}
e^{i\mathbf{k}\cdot\mathbf{x}}
+
\text{h.c.}
\right]\,,
\nonumber
\end{equation}
where $e_{ij}(\mathbf{k},s)$ are the polarization tensors,
while the amplitudes $h_{s,k}$
satisfy the same equation of motion as a massless test
scalar field, i.e.
\begin{equation}
\ddot h_{s, k} + 3 H \dot{h}_{s, k} + \frac{k^2}{a^2} h_{s, k} = 0 \,.
\label{eq_motion-pol-ten}
\end{equation}
We then obtain
\begin{equation}
\langle
\frac{\partial^i\partial_k}{\nabla^2}
\left(
h^{km}
\dot{h}_{mi}
\right)
\rangle
=
-
\langle
h_{ij}
\dot{h}^{ij}
\rangle\,,
\end{equation}
and Eq. (\ref{st2}) reduces to
\begin{equation}
\langle
s_t
\rangle
=
-
\frac{7}{64H}
\langle
h_{ij}
\dot{h}^{ij}
\rangle.
\label{stfinal}
\end{equation}
Now, by averaging Eq. (\ref{laplacianobeta}), and using Eqs. (\ref{psidott}), (\ref{qtfinal}), and (\ref{stfinal}), we find
\begin{equation}
\frac{\dot{\phi}^{\,2}}{H M_{Pl}^2}
\langle
\dot{\bar{\psi}}^{(2)}_t
\rangle
=
\left(
-
\frac{H}{4}
+
\frac{7}{32}\frac{V}{H M_{Pl}^2}
\right)
\langle
h_{ij}
\dot{h}^{ij}
\rangle \,,
\end{equation}
and at leading order in $\varepsilon$ we obtain that
\begin{equation}
\langle
\dot{\bar{\psi}}^{(2)}_t
\rangle
=
\frac{13}{64\epsilon}
\langle
h_{ij}
\dot{h}^{ij}
\rangle\,,
\end{equation}
where, by using the fact that the polarization amplitude $h_{s,k}$ satisfies the same equation of motion as a massless test scalar field (see Eq.(\ref{eq_motion-pol-ten})) and following \cite{Finelli:2004bm} (see also \cite{Starobinsky:1979ty,Marozzi:2014xma,Finelli:2008zg}), one obtains
\begin{equation}
\langle\dot h_{ij} h^{ij} \rangle
=\frac{4}{\pi^2}\frac{H^3}{M_{Pl}^2}.
\end{equation}

\section{Quantitative analysis of fine-tuning}
\label{tuning}

As systematically analyzed in \cite{Gow:2020bzo,Cole:2023wyx}, in single-field USR models, where backreaction is not considered, the PBHs abundance is extremely sensitive to the details of the dynamics near the quasi-flat region of the potential. This large sensitivity, typically identified as a fine-tuning problem  \cite{Cole:2023wyx, Frolovsky:2023hqd,Geller:2022nkr}, is usually quantified through the logarithmic measure
\begin{equation}
F_O(P)=\left|\frac{d\ln O}{d\ln P}\right|,
\label{criteriotuning}
\end{equation}
where $O$ denotes the observable under consideration, typically the amplitude of the scalar power-spectrum peak or the PBHs abundance, while $P$ is a model parameter on which that observable depends. In the following, we use this criterion to analyze the sensitivity of the entire spectrum to the parameter $\phi_d$, using the relation
\begin{equation}
F_{\Delta_{\mathcal R}^{2}}(k;\phi_d)
=\left|\frac{
\ln \Delta_{\mathcal R}^{2}(k;\phi_{d2})
-
\ln \Delta_{\mathcal R}^{2}(k;\phi_{d1})
}{
\ln \phi_{d2}
-
\ln \phi_{d1}
}
\right|\,.
\end{equation}
\begin{figure}[h]
    \centering    \includegraphics[width=\columnwidth]{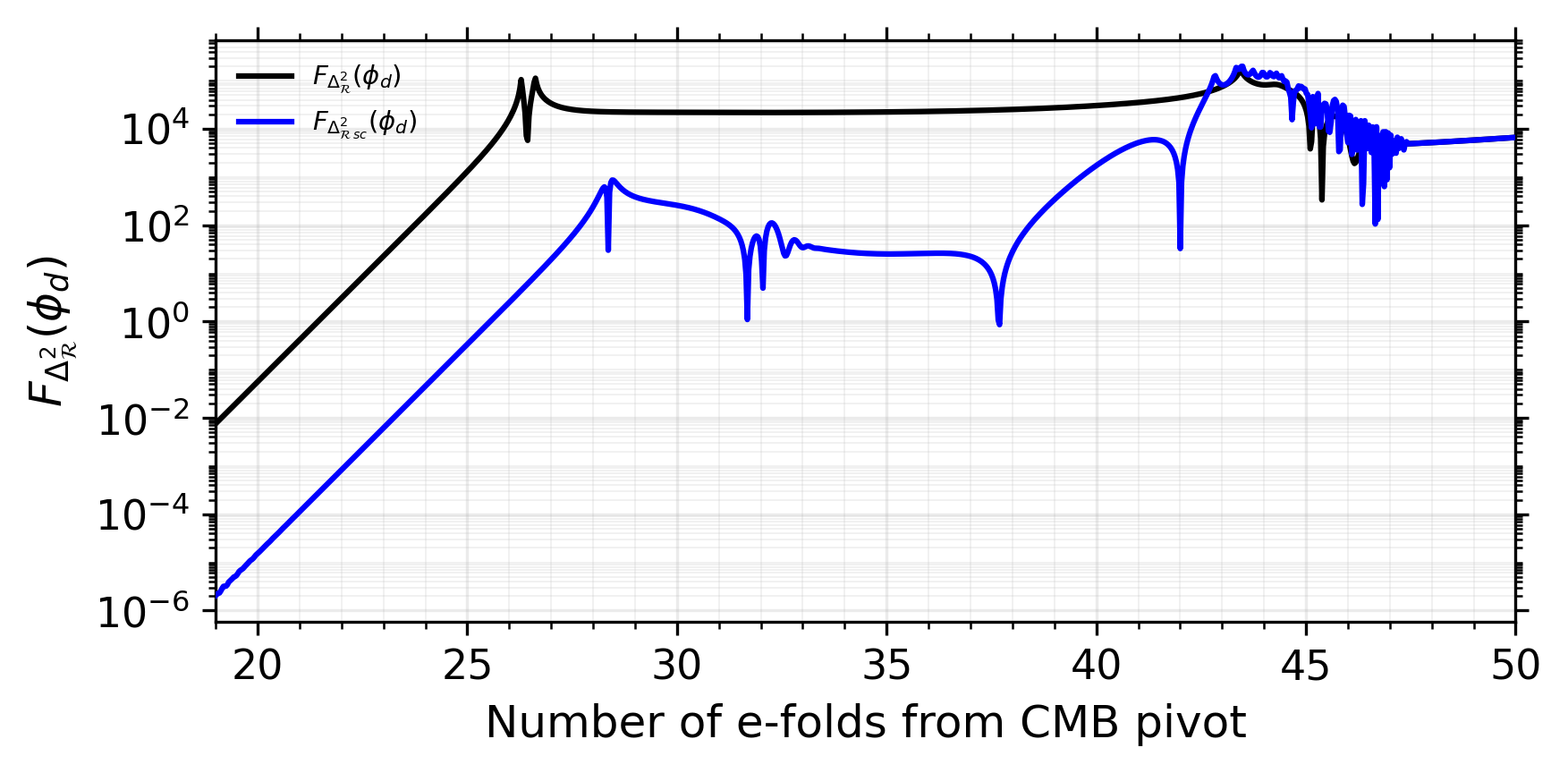}
    \caption{The black curve is the fine-tuning measure $F_{\Delta_{\mathcal R}^2}(k,\phi_d)$ for the background, while the blue curve is
    $F_{\Delta_{\mathcal R\,sc}^2}(k, \phi_d)$ for the self consistent case. They are plotted in function of the e-folds for which the mode $k$ exit the horizon.}
    \label{fig:tuning}
\end{figure}

As shown in Fig. \ref{fig:tuning}, all modes whose super-Hubble evolution is influenced by tensor backreaction exhibit reduced parametric sensitivity. For these modes, the growth of the non-constant contribution is governed by the self-consistent dynamics rather than solely by the parameters controlling the quasi-flat region of the potential. As a result, the accumulated amplification becomes less sensitive to variations of these parameters. This effect extends to modes that exit the horizon before the onset of the USR phase, because their subsequent super-Hubble evolution overlaps with the interval in which backreaction modifies the dynamics. Conversely, modes whose relevant evolution occurs after the self-consistent dynamics has converged back to the uncorrected background recover the parametric sensitivity of the background solution.
This analysis suggests that tensor backreaction can thus reduce the fine-tuning of transient USR scenarios, although, in the cases considered, the associated PBHs production is strongly suppressed.

\bibliographystyle{elsarticle-num}
\bibliography{bibPLB}

\end{document}